# Theoretical Simulation and Experiment Investigation of X-ray transmission characteristics though Square Polycapillary Slice Lens with quadratic curve

Mo Zhou[1,2,3], Kai Pan[1,2,3,6], Tian-Cheng Yi[4,5], Jian-Rong Zhou[4,5], Xue-Peng Sun[1,3], Song-Ling Wang[4,5], Xing-Fen Jiang[4,5], Bin Zhou[4,5], Bo-Wen Jiang[6], Tian-Xi Sun[1,2,3], Zhi-Guo Liu[1,2,3] and Yu-De Li[1,2,3]

[1]*College of Nuclear Science and Technology, Beijing Normal University, Beijing 100875, China*
[2]*Beijing Key Laboratory of Applied Optics, Beijing 100875, China*
[3]*Key Laboratory of Beam Technology, Ministry of Education, Beijing 100875, China*
[4]*Spallation Neutron Source Science Center, Dongguan 523803, China*
[5]*Institute of High Energy Physics, Chinese Academy of Sciences, Beijing 100049, China*
[6]*North Night Vision Technology (Nanjing) Research Institute Co.,Ltd, Nanjing 211106,China*

**Abstract**

The x-ray polycapillary lens is an optical device with good optic performance. Similar to the traditional X-ray polycapillary lens, square polycapillary slice lens was regulated on X-ray based on the full reflection principle of X-ray in the capillaries surfaces. According to its geometrical structure model and the X-ray tracing principle, a set of X-ray transmission procedures was established. A complete square polycapillary slice lens with quadratic curve was produced and the optical performance was tested

## 1. Introduction

X-ray is electromagnetic wave with high frequency and short wavelength. With the rapid development of the X-ray Analysis Technology, however, people put forward higher requirements for the cross-section and flux-density of X-ray. The x-ray polycapillary lens is an optical device with good adjustment ability of X-ray direction, focal spot size, flux density and X-ray divergence[1-6]. Hence, many potential applications have been triggered by the development of x-ray polycapillary optics, including x-ray diffraction analysis, high resolution coded-aperture imaging, x-ray lithography, x-ray fluorescence analysis, medical imaging and astronomy[7-13].

Traditional X-ray polycapillary lenses can transmit X-rays efficiently based on the principle of total reflection. In recent years, a large number of computer applications have been used to predict the transmission properties of a polycapillary x-ray lens and describe its optical properties. Liu's model uses the SHADOW ray-tracing method to simulate X-ray propagation through a polycapillary X-ray lens[14], [15]. Peng's model uses a numerical calculation to simulate X-ray transmission through a polycapillary parallel X-ray lens[16]. Wang built a model of a variable diameter multi-capillary parallel X-ray lens in MATLAB. Chi simulated the transmission characteristics of the polycapillary lens in Geant4. However, in all the above simulations the capillaries are round, which will produce a lot of shortcomings. X-rays will diffuse at the focus during multiple reflections inside the round capillaries. A triangular gap will be generated between the round capillaries, which will affect the space occupancy ratio and transmission efficiency of the lens. Since there is no clear object-image correspondence, the X-ray polycapillary with round cannot be used as an imaging optical device.

In this paper, similar to the traditional X-ray polycapillary lens, square polycapillary slice lens is regulated on X-ray based on the full reflection principle of X-ray in the capillaries surfaces. According to its geometrical structure model and the X-ray tracing principle, a set of X-ray transmission procedures was established. Finally, a complete square polycapillary slice lens with quadratic curve was produced and the optical performance was tested. The experimental results were compared with the simulation results.

## 2. Simulation model
### A. Geometry Model of the SPSL with quadratic curve

The three-dimensional schematic diagram of the SPSL with quadratic curve is shown in Fig. 1(a). The shape is a quadratic curve in the axial direction, with the multi-channel internal structure. The width of the square capillary radius gradually decreases from the entrance to the exit. Fig. 1(b) shows the lens' model in the Cartesian coordinates. The center of the lens is the origin of the coordinate system, and the incident direction of the parallel X-ray source is the positive direction of the Z-axis.

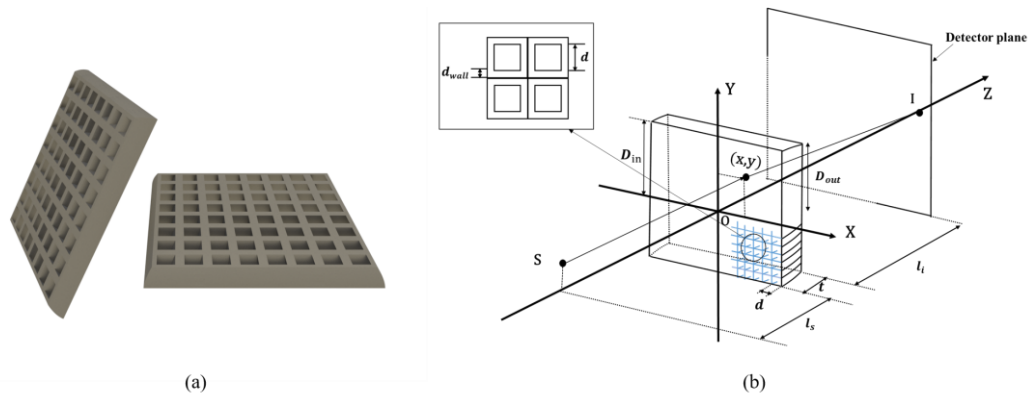

Fig. 1.(a) Three-dimensional schematic diagram of SPSL with quadratic curve. (b) Geometric model of SPSL with quadratic curve.

The travel path of X-rays in this SPSL is the same as that of other conventional lenses in the axial direction. In one-dimensional direction, the contour curve equation of the SPSL with quadratic curve is:

$$x = a_0 z^2 + b_0 z + D_{in} \quad (1)$$

where $a_0$ is the first-order coefficient, $b_0$ is the quadratic coefficient and $D_{in}$ is the half-width of the lens' entrance. The X-ray are deflected by multiple reflections in curving multi-channel internal structure and forms a X-ray focus on the outside.

**B. Ray-path Simulation in the SPSL with quadratic curve**

Fig. 2 shows the ray-path simulation of a single reflection, in one-dimensional direction. O is the origin of the X-Z coordinate system. $AO_1$ is the height of incident parallel X-ray. The coordinate of point A is $(z_A, x_A)$. The line $AA_1$ is the tangent at point A. From the geometric relationship and the Small Angle Approximation, the coordinate of point B is:

$$\left( \frac{3a_0 z_A^2 + b_0 z_A - D_{in}}{4a_0 z_A + 2b_0}, 0 \right) \quad (2)$$

In one-dimensional direction, it can be seen that the intersection with the Z-axis is mainly related to the incident X-ray coordinate and the shape curve parameters of the lens. In two-dimensional directions, the focus at the focal plane is still cross-shaped.

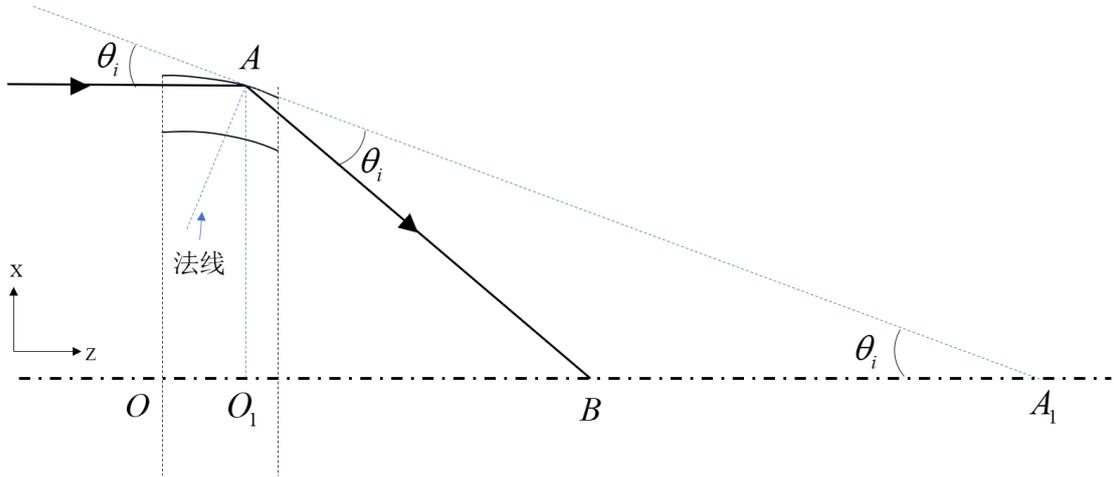

Fig. 2. SPSL with quadratic curve focusing principle (single reflection)

Fig. 3 shows the ray-path simulation of twice reflections, in one-dimensional direction. A and B are the intersection points of the first and second total reflections in lens, respectively. $AA_1$ and $BB_1$ are the tangents at point A and B. According to the geometric relationship:

$$\angle BCO_2 = 2(\theta_2 - \theta_{i1}) \quad (3)$$

the coordinate of point C is:

$$\left(0, z_A + z_B + \frac{a_0 z_B^2 + b_0 z_B + D_{in}}{4a_0(z_A - z_B)}\right) \quad (4)$$

It can be seen that in twice total reflections the intersection with the Z-axis is mainly related to the incident X-ray coordinate and the shape curve parameters of the lens. In two-dimensional directions, the focus at the focal plane is still cross-shaped. Due to the difference in the incident X-ray and the profile curve of the lens, the focus has a focal depth and the cross shape has a width.

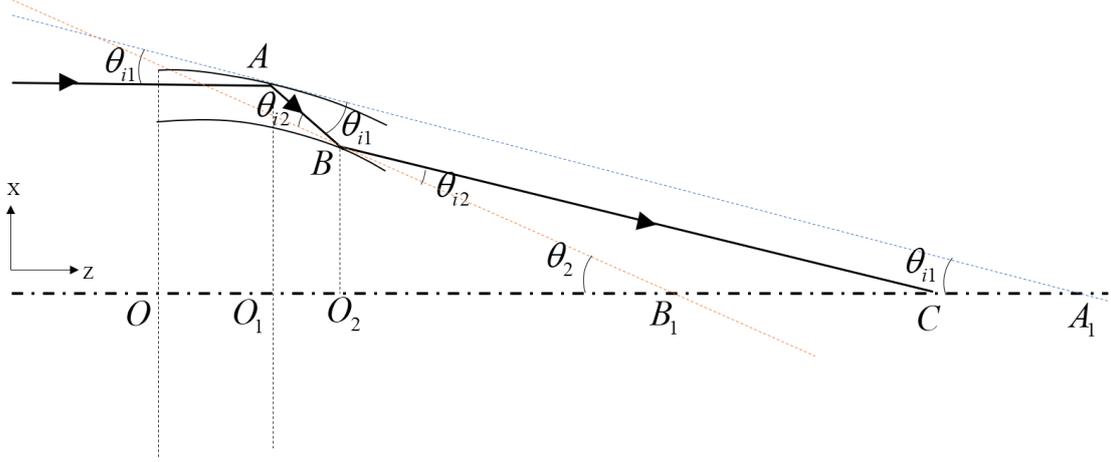

Fig. 3. SPSL with quadratic curve focusing principle (twice reflections)

### C. Modification of Calculation Model

Each capillary has its own different curvature both horizontally and vertically in SPSL with quadratic curve, so the calculation method of reflection width needs to be re-established. When the Z-axis coordinates of the parallel X-ray points are the same, as shown in Fig. 4, the minimum value of the grazing incidence angle can be considered as monotonically increasing. In the positive direction of the X-axis, there are a total of n layers capillaries and $\theta_{ori}$ is the minimum value of the initial grazing incidence angle in the m layer. The incident height ranges of the parallel X-ray though the i capillary is:

$$h_i \in \begin{cases} [\frac{1}{2}d_{SP} + (n-2)d_{SP}, \frac{1}{2}d_{SP} + (n-1)d_{SP}], i = 2,3,4...n \\ [0, \frac{1}{2}d_{SP}], i = 1 \end{cases} \quad (5)$$

where $d_{SP}$ is the capillary entrance width.

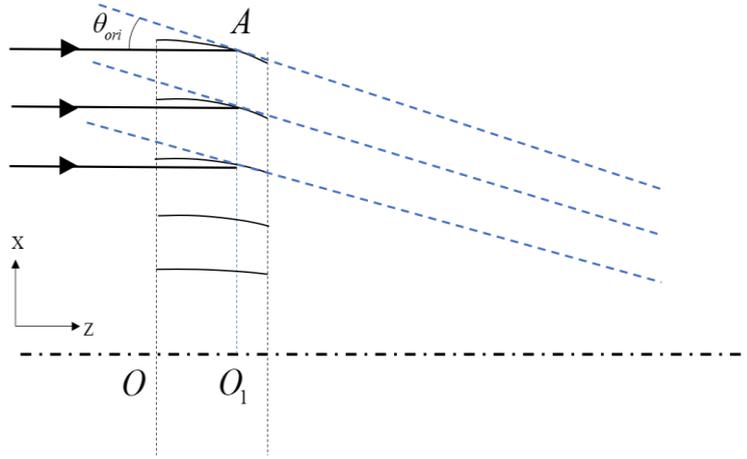

Fig. 4. Schematic diagram of minimum initial grazing incidence angles in one-dimensional direction.

When the shape parameters are determined, the width of n reflections is mainly related to the minimum value $\delta_n(\theta_{ori})$ of the initial grazing incidence angle. Similar to the ray tracing method, the calculation of the n reflections width is to determine the number of points where the X-ray collides with the capillary walls. The straight line equation of the incident X-ray though the i capillary is:

$$x_{ij} = \begin{cases} h_{i-1} + j d_{SP}/n_h, & j = 2,3,4...n_h \\ d_{SP}/2n_h, & j = 1 \end{cases} \quad (6)$$

where $n_h$ is the number of evenly divided points within the range of $h_i$. In the one-dimensional direction, the upper wall equation of the capillary is:

$$x_i = \frac{i-0.5}{n+0.5} a_0 z^2 + \frac{i-0.5}{n+0.5} b_0 z + \frac{i-0.5}{n+0.5} D_{in} \quad (7)$$

the lower wall equation of the capillary is:

$$x_i = \begin{cases} \dfrac{i-1.5}{n+0.5} a_0 z^2 + \dfrac{i-1.5}{n+0.5} b_0 z + \dfrac{i-1.5}{n+0.5} D_{in}, & i = 2,3,4...n \\ \dfrac{-i+0.5}{n+0.5} a_0 z^2 + \dfrac{i-1.5}{n+0.5} b_0 z + \dfrac{-i+1.5}{n+0.5} D_{in}, & i = 1 \end{cases} \quad (8)$$

In Fig. 5, the four colors lines correspond to the transmission of the four types of X-ray in the capillary respectively. Because of the increasing of the capillaries curvature in the axial coordinate, the parallel incident X-ray is only reflected multiple times on one side wall. This kind of reflection named "Garland reflection"【】. When the total reflection condition is satisfied, the thickness of the lens, the entrance width of a single capillary and the layers increasing will increase the number of the X-ray reflections in the capillaries.

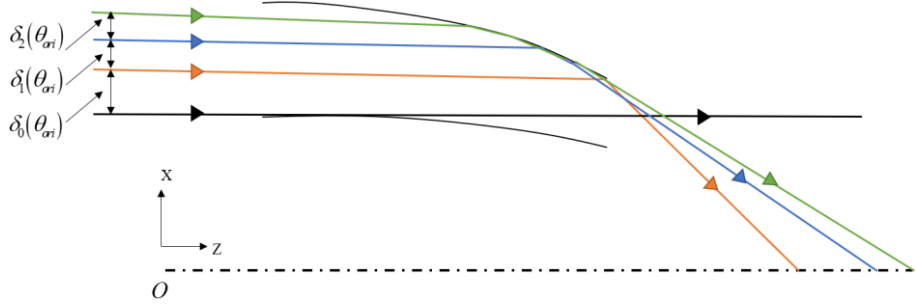

Fig. 5. n times reflection width of SPSL with quadratic curve.

Table 1 shown the parameters of the SPSL with quadratic curve. The calculated relationship between the lens and the first grazing incidence angle was shown in Fig. 6. In this lens, the X-rays were reflected in the capillaries up to 4 times. With the increase of the initial grazing incidence angle, the X-ray gradually changed from low-order reflection to high-order reflection.

Table 1. The parameters of the SPSL with quadratic curve.

| $D_{in}(mm)$ | $d_{SP}(\mu m)$ | quadratic coefficient | first-order coefficient | capillary layers |
|---|---|---|---|---|
| 10.01 | 20 | -0.00055696 | -0.00373 | 500 |

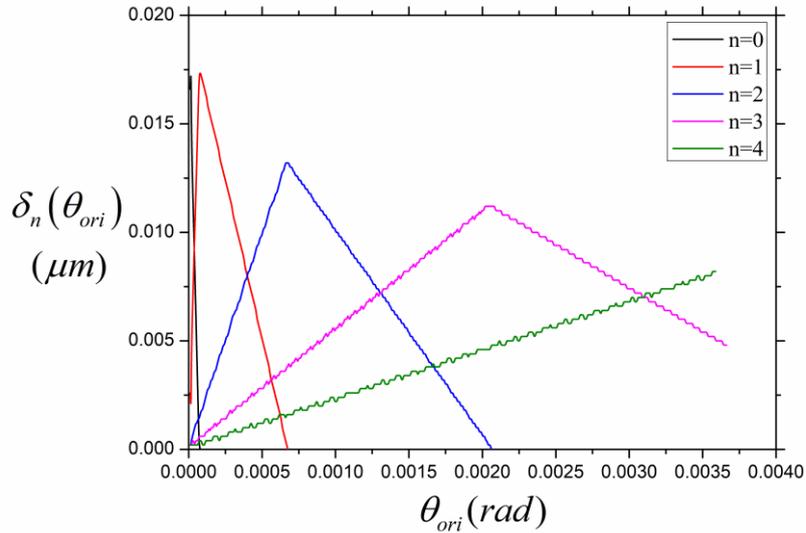

Fig. 6 The relationship between the SPSL with quadratic curve and the first grazing incidence angle.

When the first grazing incidence angle is fixed, let the maximum number of reflections be $n_{SP}$, the maximum reflection width be $\delta_{i_{SP}}(\theta_{ori})$, the number of reflections with the maximum reflection width be $i_{SP}$ and $i_{SP}$ to $n_{SP}$ reflected

X-ray be focused on the Z-axis. X-Ray from 1 to $i_{SP}-1$ reflections are diverging at the Z-axial focus. In two-dimensional direction, as shown in Fig. 7, due to the "corner-square effect", the X-ray in the A area formed a central square focus; in the B and C areas can only be focused in one-dimensional direction; in D area formed the background. Since the propagation in the capillary is a kind of "garland" reflection, the boundaries of A to D areas were determined by the number of reflections $i_{SP}$ with the maximum reflection width. Since the secondary reflection width was the largest, the average intensity of X-ray at the focal spot was the largest.

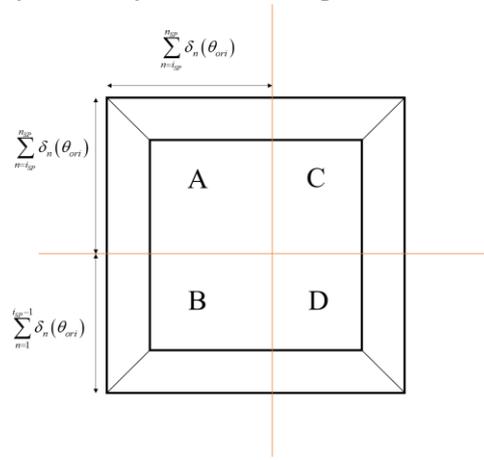

Fig. 7 The "corner-square effect" of SPSL with quadratic curve.

In order to better study the X-ray regulation performance of the SPSL with quadratic curve, the corresponding ray tracing model must be established. In the lateral direction, assuming that the device has a total of $(2n+1)\times(2n+1)$ capillaries, its shape curve equation can be expressed as:

$$f(z) = a_0 z^2 + b_0 z + D_{in} \quad (9)$$

For a single capillary, the upper, lower, left and right inner walls equations are:

$$\begin{cases} up: \dfrac{2i+\eta_{SP}}{2n_{SP}+1}\left(a_0 z^2 + b_0 z + D_{in}\right) = y \\ down: \dfrac{2i-\eta_{SP}}{2n_{SP}+1}\left(a_0 z^2 + b_0 z + D_{in}\right) = y \\ left: \dfrac{2i-\eta_{SP}}{2n_{SP}+1}\left(a_0 z^2 + b_0 z + D_{in}\right) = x \\ right: \dfrac{2i+\eta_{SP}}{2n_{SP}+1}\left(a_0 z^2 + b_0 z + D_{in}\right) = x \end{cases} \quad (10)$$

where $\eta_{SP}$ is the space occupancy ratio, This formula is a basic mathematical description, and the structural defects of the SPSL with quadratic curve are not considered.

### D. Performance Parameters
#### a. Local reflection efficiency

In one dimension, the ratio of the beam reflection width to the exit width is approximately equal to the ratio of the capillary entrance width to the exit width. The local reflection efficiency of SPSL with quadratic curve is:

$$\varepsilon_{n_x,n_y}(\theta_x,\theta_y) = \frac{d_{SP}^2}{d_{SPout}^4} \sum_{nx=i_{SP}}^{n_i} \prod_{n=1}^{nx} R(\theta_x)\delta_{n_x}(\theta_{ori\_x}) \sum_{ny=j_{SP}}^{n_j} \prod_{n=1}^{ny} R(\theta_y)\delta_{n_y}(\theta_{ori\_y}) \quad (11)$$

#### b. X-ray intensity distribution on the focal plane

Due to the curvature change of the upper and lower wall surfaces in a single capillary, the outgoing X-rays have a certain divergence. The background at the focal plane is mainly caused by the divergence of the X-rays. Incident X-rays that converges in both dimensions are called converged-converged rays (CC rays). The X-rays that converges in one direction and diverges in another direction are called converged-diverged rays (CD rays) or diverged-converged rays (DC rays). The X-rays that diverges in both dimensions are called diverged-diverged rays (DD rays). The CC rays made up the central focal, the CD or DC rays made up the cross focal lines, and the DD rays made up the background light at the focal plane. In fact, similar to the traditional polycapillary lens, because the curvature of each capillary of the lens is different, the divergence angle of the outgoing X-ray is different. Therefore, the SPSL with quadratic curve does not have a strict one-to-one correspondence between objects and images. The intensity distributions of the 3 kinds of X-rays at the focal plane are:

$$I_{cc}(x,y) = \frac{N}{4D_{in}^2} \sum_{i=i_{SP}}^{n_{SP}} \int_{X_c(\theta_{ori})} \prod R(\theta_x) d\theta_{ori\_x} \sum_{j=i_{SP}}^{n_{SP}} \int_{Y_c(\theta_{ori})} \prod R(\theta_y) d\theta_{ori\_y}$$

$$I_{cd}(x,y) = \frac{N}{4D_{in}^2} \sum_{i=i_{SP}}^{n_{SP}} \int_{X_c(\theta_{ori})} \prod R(\theta_x) d\theta_{ori\_x} \sum_{j=1}^{i_{SP}-1} \int_{Y_d(\theta_{ori})} \prod R(\theta_y) d\theta_{ori\_y}$$

$$I_{dd}(x,y) = \frac{N}{4D_{in}^2} \sum_{i=1}^{i_{SP}-1} \int_{X_d(\theta_{ori})} \prod R(\theta_x) d\theta_{ori\_x} \sum_{j=1}^{i_{SP}-1} \int_{Y_d(\theta_{ori})} \prod R(\theta_y) d\theta_{ori\_y} \quad (12)$$

#### c. Collection efficiency of incident X-ray

Before calculating the X-ray collection efficiency of a SPSL with quadratic curve, a one-dimensional collection efficiency function needs to be determined. The one-dimensional convergent collection efficiency and divergent collection efficiency are:

$$F_c(\theta_{ori}) = \frac{\int_{\theta_{ori}} \Pi R(\theta) \sum_{i=i_{SP}}^{n_{SP}} \delta_n(\theta_{ori})}{f}$$

$$F_d(\theta_{ori}) = \frac{\int_{\theta_{ori}} \Pi R(\theta) \sum_{i=1}^{i_{SP}-1} \delta_n(\theta_{ori})}{f}$$

(3-1)

The collection efficiencies for three types of X-rays are:

$$\Omega_{cc} = [F_c(\theta_{ori})]^2$$
$$\Omega_{cd} = F_c(\theta_{ori}) \Box F_d(\theta_{ori})$$
$$\Omega_{dd} = [F_d(\theta_{ori})]^2$$

(3-2)

### d. Peak-to-background ratio at the focal plane

the SPSL with quadratic curve, as an X-ray focusing lens, can be used as a focusing element in a magnified imaging system. Therefore, the influence of the peak-to-background ratio at the focal plane on the imaging resolution must be considered. The peak-to-back ratio at the focal plane can be calculated from the ratio of the CC rays to the CD rays average intensities at the focal plane.

## 3. Simulation result
### A. Local reflection efficiency

Fig. 8 was a two-dimensional distribution diagram of the local reflection efficiency of the SPSL with quadratic curve when the incident X-ray is 6.4 keV. The local reflectivity model adopted a double-factor model. The local reflection efficiency in the central area of the lens was the highest, and the two-dimensional direction distribution of local efficiency is like a cross. When the first grazing incidence angle increased, the local reflection efficiency in the one-dimensional direction exhibited a fluctuating attenuation trend. Since X-rays formed higher-order reflections when the first grazing incidence angle increased, the total reflection width for focusing drops abruptly as the $i_{SP}-1$ reflections move toward $i_{SP}$ reflections. This created a step situation for the x-rays in the two-dimensional distribution.

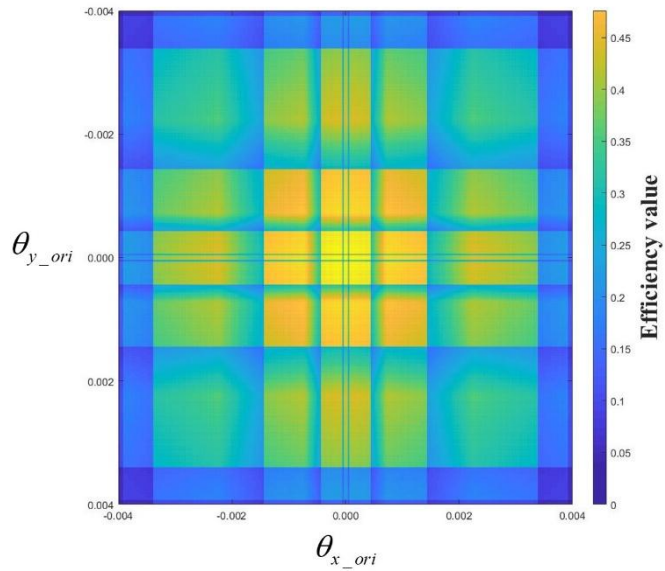

Fig. 8 The two-dimensional distribution diagram of the local reflection efficiency of the SPSL with quadratic curve.

### B. X-ray intensity distribution on the focal plane

Fig. 9(a) shown the X-ray intensity distribution of the SPSL with quadratic curve at the focal plane calculated by formula (4-187) when the X-ray energy is 6.4keV and the number of photons is N=4000000. The X-ray intensity distribution is like a cross, and the width of the cross focal line gradually decreases from the center to the edge. Fig. 9(b) shown the one-dimensional distribution of X-ray intensity on the X-axis. In the central, the CC rays had sharp "tower" distributions, and the full width of the focal spot is 110 $\mu m$.

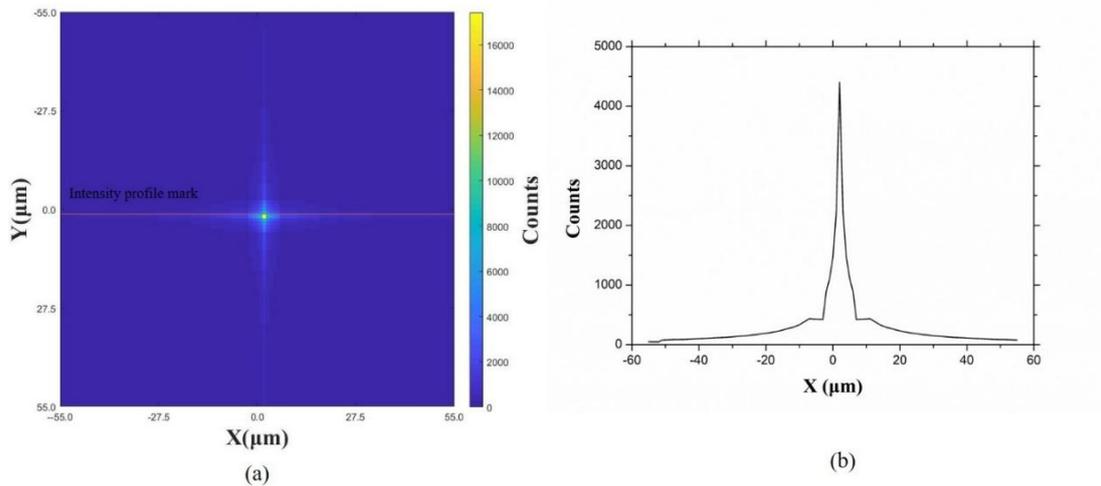

Fig. 9(a) Two-dimensional X-ray intensity distribution and (b) one-dimensional X-ray intensity distribution of the SPSL with quadratic curve at the focal plane.

### C. Collection efficiency of incident X-ray

Fig. 10 showed the collection efficiency of the lens for different types of rays as a function of first grazing incidence angle. It can be seen that the overall collection efficiency of the lens for X-rays decreased as the first grazing incidence angle increases. The collection efficiency of the device for CC rays fluctuated and decreased with the increase of the first grazing incidence angle, while the collection efficiency for CD rays and DD rays increased with the increase of the first grazing incidence angle. This shown that with the increase of the overall shape curvature of the lens, the overall X-ray collection efficiency gradually decreases and the interference of the focal spot at the focal plane by the background light increases.

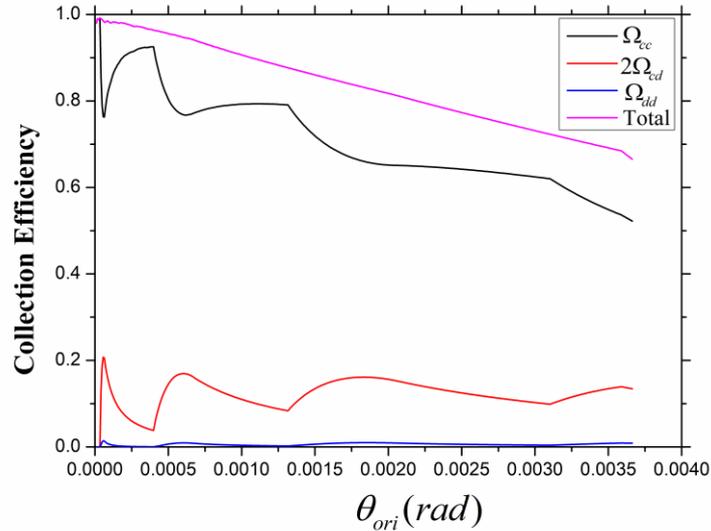

Fig. 10 The relationship between the X-ray collection efficiency of the SPSL with quadratic curve and the first grazing incidence angle.

### D. Peak-to-background ratio on the focal plane

Fig. 11 showed the curve of the peak-to-background ratio of the focal plane as a function of the first grazing incidence angle. As the first grazing incidence angle increases, the peak-to-background ratio decreases and the influence of the one-dimensional focal line on the central focal spot increases.

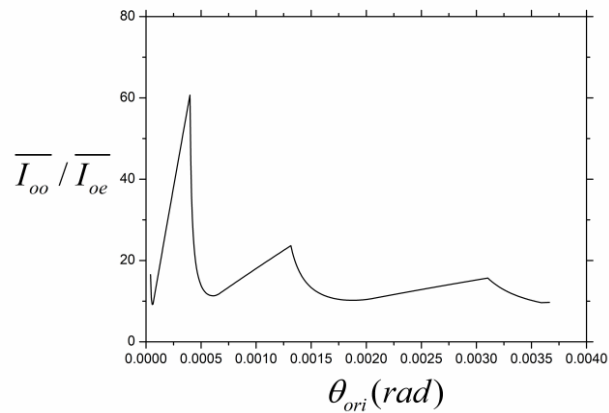

Fig. 11 The curve of the peak-to-background ratio of the focal plane as a function of the first grazing incidence angle.

## 4. Experiment result

Fig. 12 is a raw glass rod. An axial scan of the diameter of the glass rod is required prior to cutting out the lens. Find the part that conforms to the quadratic curve shape from the scanned image and cut it.

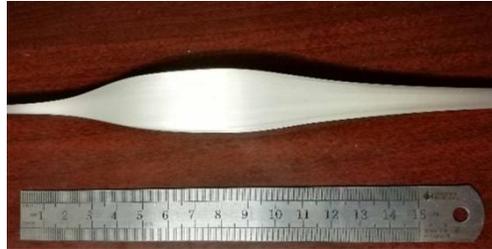

Fig. 12 A raw glass rod.

Fig. 13 is a data graph of a diameter scan of the glass rod. The red and green curves in the figure indicate the scanning results in the horizontal and vertical directions, respectively. The blue curve is the appropriate interception position. It can be seen from the difference between the curves that, in fact, the glass rod cannot guarantee a standard square cross-section due to the manufacturing process error. A batch of SPSLs with quadratic curve was obtained by cutting, grinding and etching processes as shown in Fig. 14.

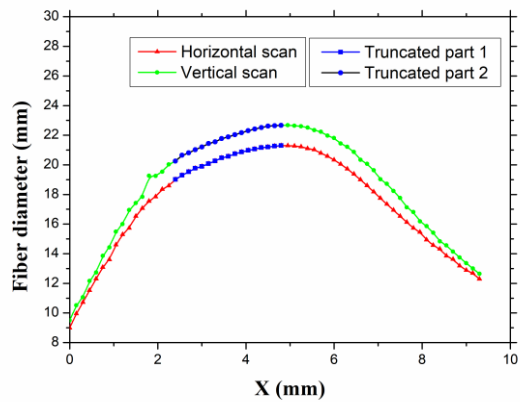

Fig. 13 A data graph of a diameter scan of the glass rod.

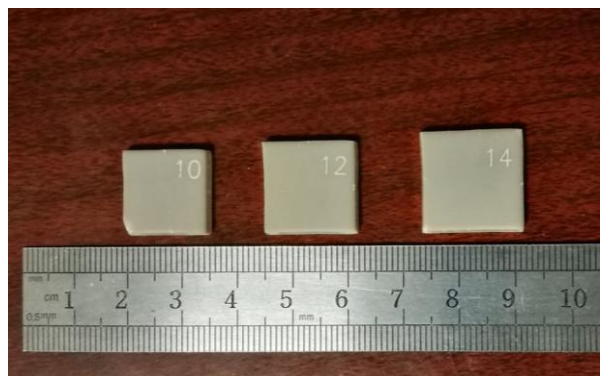

Fig. 14 a batch of SPSLs with quadratic curve.

Through screening, the best No. 18 SPSL with quadratic curve was selected for further optical performance test. The No. 18 SPSL with quadratic curve is shown in

Fig. 15, and its specific parameters are shown in Table 2.

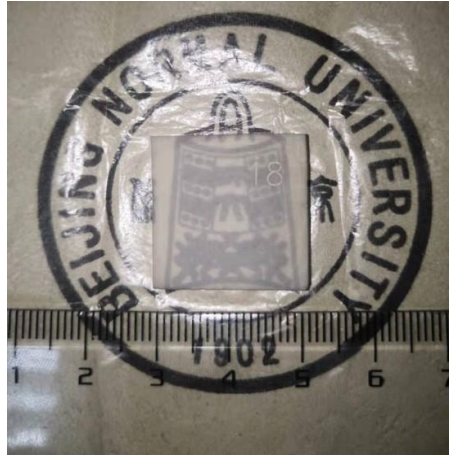

Fig. 15 The No. 18 SPSL with quadratic curve

Table 2. The No. 18 SPSL with quadratic curve Optical Parameters

|  | $t_{SP}$ (mm) | $D_{in}$ (mm) | $D_{out}$ (mm) | The number of capillaries | Space Occupancy Ratio |
|---|---|---|---|---|---|
| No. 18 | 3 | 10.71 | 10.50 | 1080×1080 | 0.8 |

Fig. 16 is a focal spot image of the No. 18 SPSL with quadratic curve. Its focal spot size is 300 $\mu m$, which is smaller than the simulation result of 500 $\mu m$. The main reason is that only the central area of the No. 18 SPSL can receive strong X-ray due to the limitation of the X-ray output area of the front-end lens. The outer capillaries transmit less X-rays due to the large change in the profile curve of slice No. 18.

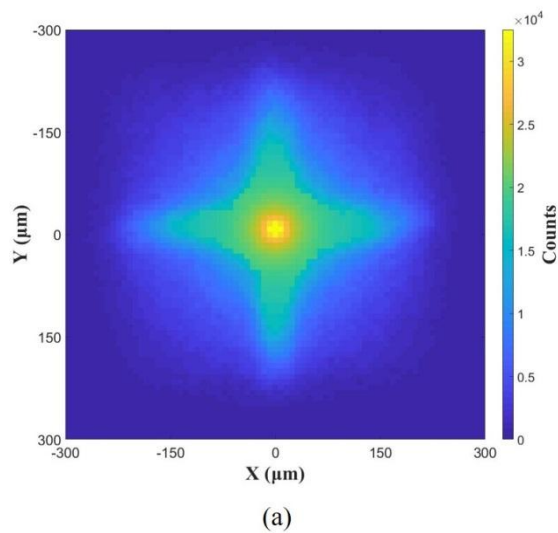

(a)

Fig. 16 A focal spot image of the No. 18 SPSL with quadratic curve.

As a polycapillary lens, overall divergence angle and transmission efficiency are important performance parameters. The simulated result of the overall divergence angle of the No. 18 SPSL is 0.079rad, and the experimental result is 0.023rad. The transmission efficiency of the No. 18 SPSL for incident X-ray is 11%, which is lower than the simulation result of 23%.

## 5. Conclusion

In this paper, a novel X-ray SPSL with quadratic curve is proposed. The X-rays transmission paths in the lens are simulated, and the simulation model for calculating the reflection width is revised. Under the condition of different first grazing incidence angles, according to the structural characteristics of the lens, the local reflection efficiency, the X-ray intensity distribution on the focal plane, the collection efficiency of incident X-ray and the peak-to-background ratio on the focal plane of the lens were calculated. The optical performance test was carried out on the No. 18 SPSL with quadratic curve, and the focusing performance such as focal spot, overall divergence angle and transmission efficiency were tested respectively. The applicability of the computational model and the tracing model is verified by the comparative analysis of the experimental results and the simulation results.

**Acknowledgment**
This work was supported by the National Natural Science Foundation of China (Grant No. 12175021), the National Natural Science Foundation of China (Grant No. 12175254) and the Guangdong Basic and Applied Basic Research Foundation (Grant No. 2019A1515110398).